# THE RELATIONSHIP BETWEEN
# INERTIAL AND GRAVITATIONAL MASS

by

Gregory W. Horndeski
2814 Calle Dulcinea
Santa Fe, NM 87505-6425
e-mail:
horndeskimath@gmail.com

March 6, 2016

**ABSTRACT**

I will argue that the inertial mass is greater than the gravitational mass for all objects which are held together primarily by gravitational forces. Thus, for celestial objects, like planets, stars and galaxies, their inertial mass is greater than their gravitational mass. The analysis used to arrive at this conclusion shows that there should, in principle, exist classical objects with non-zero inertial mass and vanishing gravitational mass. Implications for Quantum Gravity and Quantum Cosmology, are briefly discussed.

As we know, the earth is held together by gravity. Thus if you are exterior to the earth, the gravitational field you experience is not the total field that the earth is capable of producing, but only what remains after the earth's "gravitational binding field" has been removed. This situation is similar to the strong force that holds nuclei together. For example, let's consider an alpha particle. The inertial mass of this particle will be less than the sum of the inertial mass of the two protons and two neutrons that constitute it, due to the negative nuclear binding energy of the alpha particle. Similarly if you add up the gravitational mass of the earth's constituents (when they were free entities), that sum will be greater than the gravitational mass of the earth, because of the mass equivalent of the negative gravitational potential binding energy. This can also be looked at from a Quantum Gravitational point of view (*see*, *e.g.*; Horndeski [1]). There we have EPs (:=elementary particles) in the earth emitting VGs (:=virtual gravitons) in arbitrary directions. Now we cannot say that such and such EPs produce VGs that hold the earth together, while these other EPs produce VGs that are responsible for the field external to the earth. Evidently, which EPs in the earth do what, is constantly changing (and some may be doing both simultaneously). Consequently the macroscopic gravitational field we measure external to the earth, is actually the time average of those VGs which escape from the earth's interior, which is less that the total number of VGs being produced. So there is no question in my mind that the inertial mass must be greater than the gravitational mass. The problems is, just how great is the effect, and when does it become a significant issue.

The idea that the gravitational potential energy of a body can reduce its gravitational mass was originally developed by Kenneth Nordtvedt, Jr., in [2]. He also proposed experiments where his "Nordtvedt Effect" involving the earth and the moon could be observed [3]. As of yet these experiments have not demonstrated any difference between the earth's or moon's inertial and gravitational mass (see [4]). More will be said about the Nordtvedt Effect latter.

I will now sketch the arguments presented by Typinski in [5] which show how to use Newtonian theory to compute the Earth's gravitational potential energy. The gravitational mass of the earth is given by (see [6])

$$M_{\oplus g} = (5.9722 \pm .0006) \times 10^{24} \text{ kg,} \qquad \text{Eq.1}$$

which includes the mass of the earth's atmosphere. In [5] Typinski shows that if we assume the earth to be a ball of uniform density then its gravitational potential energy, $U_{unif}(M_{\oplus g})$, would be given by

$$U_{unif}(M_{\oplus g}) = -3 \, G \, M_{\oplus g}^2 / (5 \, R_\oplus), \qquad \text{Eq.2}$$

where $G = (6.67408 \pm .00031) \times 10^{-11} \text{ nt m}^2/\text{kg}^2$ is the Newtonian gravitational constant, and $R_\oplus$ is the radius of the earth. This leads to a value of

$$U_{unif}(M_{\oplus g}) = -2.242 \times 10^{32} \text{ joules,}$$

and hence

$$-U_{unif}(M_{\oplus g}) / c^2 = 2.495 \times 10^{15} \text{ kg.} \qquad \text{Eq.3}$$

If we approximate the earth as being built up from nested concentric shells, each of which is of different uniform density, then Typinski has shown that the gravitational potential energy is

$$U_{variable}(M_{\oplus g}) = -1.711 \times 10^{32} \text{ joules,}$$

with a corresponding mass of

$$-U_{variable}(M_{\oplus g}) / c^2 = 1.904 \times 10^{15} \text{ kg,} \qquad \text{Eq.4}$$

which is about 24% smaller than the mass given in Eq.3. So if we take the earth's inertial mass to be given by

$$M_{\oplus i} = M_{\oplus g} + (-\text{Earth's gravitational potential energy} / c^2),$$

then we can use either Eq.3 or Eq.4, along with Eq.1 to deduce that

$$(M_{\oplus i} - M_{\oplus g}) / M_{\oplus i} \text{ is of order } 10^{-10}. \qquad \text{Eq.5}$$

Typinski does a similar calculation for the Moon and shows that the ratio corresponding to Eq.5 for the Moon is of order $10^{-11}$. He then goes on to state that: "If the Nordtvedt effect operates at its theoretical maximum, the Moon's orbit should be elongated along the Earth-Sun radial by an amount on the order of ten meters." This effect has not been observed, [4].

My contention is that the ratio considered in Eq.5 is too large by several orders of magnitude. To see why let's consider the hydrogen atom. The gravitational potential energy for that system is approximately given by

$$U(H) = - G m_p m_e / \langle r \rangle, \qquad \text{Eq.6}$$

where $m_p$ is the inertial mass of the proton, $m_e$ is the inertial mass of the electron, and $\langle r \rangle$ is the expected radial distance of the electron from the proton in the ground state of the hydrogen atom, which is $\langle r \rangle = 1.5 \times$ Bohr Radius $= 7.9376581 \times 10^{-11}$ m. Using the usual values for $m_p$ and $m_e$ in Eq.6 we find

$$U(H) = -1.28110812 \times 10^{-57} \text{ joules}. \qquad \text{Eq.7}$$

This corresponds to a mass $U(H)/c^2 = -1.425425 \times 10^{-74}$ kg. Thus the gravitational mass of the hydrogen atom in its ground state would be its usual inertial mass of $1.6727 \times 10^{-27}$kg, less $1.425425 \times 10^{-74}$kg. It is because of minuscule differences like this that Eötvös-type experiments have such a hard time finding a difference between inertial and gravitational mass for anything smaller than celestial objects. However, there should be a difference between inertial and gravitational mass for all objects except EPs. Now what if we approximate the hydrogen atom as a ball of uniform density. Then we can use the obvious modification of Eq.2 to conclude that the gravitational potential energy of the hydrogen atom viewed as a ball, $U_{ball}(H)$, is given by

$$U_{ball}(H) = -1.411342 \times 10^{-54} \text{ joules},$$

which is about 3 orders of magnitude larger than the value given in Eq.7. This leads me to suspect that the values given for the earth's gravitational potential energy calculated by Typinski using a continuum model of the earth should be too high by at least 3 orders of magnitude, which would give life to the Nordtvert Effect. Exactly how one could go about doing some sort of discrete calculation of the earth's gravitational potential energy is unclear to me. But rather than do that, why not turn our attention to neutron stars and pulsars where a continuum model of the mass distribution is much more plausible. *E.g.,* in [7] Champion, *et al.,* present an example of a 1.74 solar-mass pulsar, PSR J1903+0327, whose gravitational potential energy corresponds to a mass of .26622 solar masses. Clearly there should be a measurable "Nordtvert-Type Effect" for this object and its companion, which is a one solar mass star. However, I have not heard of anything done on this problem. Although some work is currently being done to check the validity of the strong equivalence principle (*see*, [8] for the definition of this notion), on a stellar triple system with pulsar PSR J0337+1715 and two white dwarfs orbitting it, which was discovered by Ransom, *et al.,* [9].

It is interesting to pursue our observation concerning the gravitational mass of the hydrogen atom a bit further. Suppose we have two identical point particles with gravitational mass m. What would their separation, $r_0$, have to be, for their gravitational potential energy, $U(m,r_0)$, to be such that

$$-U(m,r_0)/c^2 = 2m \text{ ?}$$

We easily find $r_0 = Gm/(2c^2)$, which is 1/4th the Schwarzschild radius of m. At this distance of separation, our previous work would imply that the gravitational mass for this system of two identical point particles of free gravitational mass m would be 0. For most elementary particles $r_0$ is much smaller than the Planck length, $L^* := (\hbar G/c^3)^{1/2}$. But what if our two identical point particles had the Planck mass, $M^* := (\hbar c/G)^{1/2}$? Then we would find that $r_0$ for that system would be given by $L^*/2$, and gravity would vanish outside of that two particle system. In the Quantum Gravity Theory that I present in [1], it is assumed that whenever two EPs come within $L^*$ of each other, the gravitational attraction between them vanishes. As a result, in this theory's model of the early universe, which I assume to be populated by particles of mass $M^*$ that are within $L^*$ of their nearest neighbors, there is no gravity. The "classical" observations we just made regarding a system of two bodies of mass $M^*$, gives credence to what I do in [1].

    The fact that the gravitational mass of a system of particles can be less than the sum of the "free" gravitational masses of its constituents, seems to violate conservation of energy. To see that this is not so, let us go back to the alpha particle. When that particle forms from two protons and two neutrons, the creation of the new particle is accompanied by the emission of gamma rays. So that the total energy of the original protons and neutrons is equal to the energy of the alpha particle and the gamma rays. Similarly when the earth forms from its constituents, the objects which will comprise the earth fall toward the system's center of gravity. As they fall, they accelerate and emit gravitational radiation and heat as they collide. The amount of radiation and heat released is equal to the gravitational potential energy which binds the earth together. So in the end the sum of the gravitational mass of the earth, and the radiation released in its formation (which equals the gravitational potential binding energy), is equal to the original gravitational mass of the earth's constituents when they were "free" entities. If we now return to the above example of two masses of mass $M^*$, we see as they fall toward each other, they gravitationally radiate away all of their gravitational mass by the time they get to within $L^*/2$ of each other. So there really is nothing

mysterious taking place when we find that the gravitational mass of a system of particles is less than the original gravitational mass of the constituent free particles.

In passing, I would like to point out that due to the work of E. Kajari, *et al*. [10], there appear to be entities on the quantum level for which the inertial and gravitational mass can differ, and the differences can be substantial.

**ACKNOWLEDGMENTS**

I would like to thank my wife, Dr. Sharon Winklhofer Horndeski, for assistance in preparing this manuscript. I also wish to thank Professors Davood Momeni, and John Wainwright, for discussions on the topics addressed here.